\documentclass[aps]{revtex4}
\usepackage{epsfig}
\usepackage{braket}

\begin{document}
\title{Entanglement and swap of quantum states in two qubits}

\author{Takaya Ikuto}
\author{Satoshi Ishizaka}
\affiliation{
Graduate School of Integrated Arts and Sciences, Hiroshima University, Higashihiroshima, 739-8521, Japan}

\date{\today}

\begin{abstract}
Suppose that two distant parties Alice and Bob share an entangled state $\rho_{AB}$,
and they want to exchange the subsystems of $\rho_{AB}$ by local operations and classical communication (LOCC).
In general, this LOCC task (i.e. the LOCC transformation of $\rho_{AB} \to V\rho_{AB} V$ with $V$ being a swap operator) is impossible deterministically,
but becomes possible probabilistically.
In this  paper, we study how the optimal probability is related to the amount of entanglement in the framework of positive partial transposed (PPT) operations,
and numerically show a remarkable class of states whose probability is the smallest among every state in two quantum bits.
\end{abstract}


\maketitle

\section{Introduction}
\label{s1}
Suppose that two distant parties Alice and Bob share an entangled state $\rho_{AB}$,
but they need the other state such that the subsystems of $\rho_{AB}$ are exchanged (i.e. $V\rho_{AB} V$ with $V$ being a swap operator). Is it possible for them to swap a given $\rho_{AB}$ by local operations and classical communication (LOCC)?
This problem, first considered by Horodecki {\it et. al.} \cite{KMP},
is quite interesting, because it is closely related to the asymmetry of quantum entanglement.
Indeed, if the entanglement contained in $\rho_{AB}$ is asymmetric and the amounts of the entanglement of $\rho_{AB}$ and $V\rho_{AB}V$ are not equal 
(in some entanglement measure), we can immediately conclude that $\rho_{AB}$ cannot be swapped by LOCC, because the amount of entanglement cannot be increased by LOCC \cite{KMP}.
In Ref \cite{KMP}, it has also been shown that the entangled states which can be swapped by LOCC can be swapped by a local unitary transformation.
This implies that the entangled state $\rho_{AB}$ whose reduced density matrices $\rho_A$ and $\rho_B$ have different eigenvalues cannot be swapped by LOCC, 
and therefore almost all entangled states cannnot be swapped by LOCC. 
However, this is the case where a deterministic LOCC transformation with a unit probability is considered.
As shown later, such states can also be generally swapped by stochastic LOCC (SLOCC), and therefore by 
examining the success probability, it becomes possible to quantitatively discuss the degree of difficulty of swapping.
For example, it can be said that the state having the smallest swapping probability is the most difficult state to swap,
and it is expected that the entanglement of the state is the most asymmetric.

In this paper, we numerically calculate the swapping probability under positive partial transposed (PPT) operations \cite{EMR,Rains01a,Plenio,Ishizaka}, which are known to contain LOCC.
Concretely, we obtain the maximal $p$ for given $\rho_{AB}$ among trace-nonincreasing PPT operations $\Lambda$ such that $\Lambda (\rho_{AB}) = p V \rho_{AB} V$.
As a result, we provide the numerical evidence for the special class of states whose swapping probability is the smallest among every state in two quantum bits (qubits).
Moreover, we discuss why the class of states exhibits the small swapping probability.

\begin{figure}[tb]
\centerline{\epsfig{file=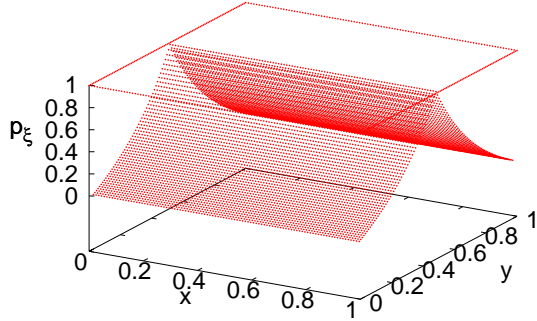, width=8.2cm}}
\vspace*{13pt}
\caption{$p_\xi$ as a function of $x$ and $y$.}
\end{figure}

\section{Swapping by SLOCC}
\label{s2}
To begin with, we show some specific examples of the swap by SLOCC. 
When $\rho_{AB}$ is not entangled,
it can be clearly swapped by SLOCC with a unit probability, because $V\rho_{AB}V$ is not entangled and can be generated by LOCC from the scratch.
When $\rho_{AB}$ is an entangled pure state $\ket{\psi}$,
it can be also swapped with a unit probability,
because $\ket{\psi}$ can always be written in the Schmidt decomposed form:
\begin{equation}{
\ket{\psi} = \sqrt{p_0}\ket{0_A0'_B} + \sqrt{p_1}\ket{1_A1'_B} +  \sqrt{p_2}\ket{2_A2'_B}\cdots ,
} \end{equation}
and hence the difference between Alice and Bob is only the local bases 
(the swap is thus achieved by a local unitary transformation exchanging the local bases).
In the other cases of entangled mixed states, the swapping probability becomes smaller than 1 in general. 
For example, let us consider the mixed state
\begin{equation}{
\xi = x\ket{\psi}\bra{\psi} + (1-x)\ket{01}\bra{01},
} \end{equation}
where $\ket{\psi} = \sqrt{y}\ket{00} + \sqrt{1-y}\ket{11}$, $0\leq x\leq 1$, and $0\leq y\leq 1$.
For later convenience, let us summarize the properties of $\xi$ here.
The concurrence of $\xi$ is given by
\begin{equation}
C_\xi = 2x\sqrt{y(1-y)} \label{concurrence},
\end{equation}
and the purity is given by
\begin{equation}
\hbox{tr}\xi^2 = \textstyle 2(x-\frac{1}{2})^2 +\frac{1}{2} \label{purity}.
\end{equation}
The reduced density matrices $\xi_A$ and $\xi_B$ are
\begin{eqnarray}
\xi_A &=& \left\{ 1-x(1-y)\right\} \ket{0}\bra{0} +x(1-y)\ket{1}\bra{1}, \\
\xi_B &=& xy\ket{0}\bra{0} + (1-xy)\ket{1}\bra{1}.
\end{eqnarray}
From the above, the absolute value of the difference of the purity of $\xi_A$ and $\xi_B$ is given by
\begin{equation}
\Delta P_\xi = \textstyle |\hbox{tr}\xi_A^2 - \hbox{tr}\xi_B^2 | = 2 | (2y-1)\{ (x-\frac{1}{2})^2-\frac{1}{4}\} | \label{dp}.
\end{equation}

Now, the optimal (maximal) swapping probability $p_\xi$ of $\xi$ by SLOCC is as follows:
\begin{itemize}
\item[(i)] For $x=0,1$ or $y=0,1$, $p_{\xi} = 1$ because $\xi$ is pure or not entangled.
\item[(ii)] For $x\neq 0,1$ and $0<y\leq \textstyle \frac{1}{2}$, $p_{\xi} = y/(1-y)$.
The swap is realized by an SLOCC $\Lambda$ as $\Lambda(\rho) = (A\otimes B)\rho(A\otimes B)^\dagger$ with
\begin{equation}
A = B = \ket{1}\bra{0} + \sqrt{\frac{\displaystyle y}{\displaystyle 1-y}}\ket{0}\bra{1}, \label{kkk}
\end{equation}
because
$\Lambda(\xi) = \frac{y}{1-y}V\xi V$.
The proof of the optimality is given by Appendix A.
It is interesting that $p_\xi$ is independent of $x$ for $0<x<1$.
\item[(iii)] For $x\neq 0,1$ and $\textstyle \frac{1}{2}\leq y<1$, it is clear that $p_\xi = (1-y)/y$ from the symmetry.
\end{itemize}
In this way, the state $\xi$ can be swapped with some finite probability even when it cannot be swapped by deterministic LOCC.
In fact, as shown in Sec. \ref{s4} below, numerical results indicate that, in the limit of $x \rightarrow 1$, $p_\xi$ is actually a lower bound for the swapping probability of \emph{any} state of the same concurrence. However, $p_\xi$ changes discontinuously to $1$ at $x=0,1$ and $y=0,1$ as shown in Fig. 1 (see Sec. \ref{s5} for further discussion of this point).

\section{Formulation by PPT operations}
\label{s3}
LOCC is the most important class of operations in entanglement manipulations, but the mathematical treatment is quite difficult.
Indeed, obtaining the optimal swapping probability for general states is a quite hard task even numerically.
Note that, the maps of having a restricted form $(A \otimes B)\rho (A \otimes B)^{\dagger}$ is sometimes considered as SLOCC (e.g. \cite{Kent}),
but in our case to obtain the optimal swapping probability,
we need to consider the most general form of SLOCC such as $\sum_i(A_i \otimes B_i)\rho (A_i \otimes B_i)^{\dagger}$ with $\sum_i A^{\dagger}_iA_i\otimes B^{\dagger}_iB_i\le I$ (though this is indeed the form of trace-nonincreasing separable operations).
In this paper, therefore, we numerically investigate it under positive partial transposed (PPT) operations \cite{EMR,Rains01a,Plenio,Ishizaka}.
The states whose density matrices are kept positive despite partial transposition are called PPT states in association with Peres' separability criterion \cite{peres},
and the operations which transform PPT states into PPT states are PPT operations. 
Since PPT operations are slightly less constrained than LOCC, the swapping probability by stochastic PPT operations is equal to or higher than the probability by SLOCC. 
Therefore, the swapping probability of unentangled states and pure states also reaches 1. 
From the results of numerical calculations, it is found that the optimal swapping probability of the mixed state $\xi$ by PPT operations is also given by Fig. 1. 

For general state $\rho_{AB}$, the swapping probability is obtained as follows.
From the so-called Jamio\l kowski isomorphism \cite{J,P,C}, let us denote $E_{{A_1}{A_2}{B_1}{B_2}}$ as the isomorphic matrix to PPT operations $\Lambda$ (i.e. $E_{A_1A_2B_1B_2} = (\Lambda_{A_1B_1} \otimes I_{A_2B_2})\Phi_{A_1A_2B_1B_2}$ with $\Phi$ being a maximally entangled state between $A_1B_1$ and $A_2B_2$). 
In order for $\Lambda$ to be PPT operations, $E_{{A_1}{A_2}{B_1}{B_2}}$ must fulfill the following conditions.

\begin{itemize}
\item[(a)]$E_{{A_1}{A_2}{B_1}{B_2}} \geq 0$, because $\Lambda$ is a completely positive map.
\item[(b)]$\textstyle E_{{A_1}{A_2}{B_1}{B_2}}^{{T_{A_1}}{T_{A_2}}} \geq 0$, because $\Lambda$ transforms PPT states into PPT states, where $T_{X}$ denotes the partial transposition with respect to the subsystem $X$.
\item[(c)]$E_{{A_2}{B_2}}\leq  \frac{\displaystyle I}{\displaystyle d}$, because $\Lambda$ is a trace-nonincreasing map,
where $E_{{A_2}{B_2}} \equiv \hbox{tr}_{{A_1}{B_1}}E_{{A_1}{A_2}{B_1}{B_2}}$.
\item[(d)]
$d^2 \hbox{tr}_{A_2B_2}E(I\otimes  \rho_{AB}^T) = pV\rho_{AB} V$
must hold when the swapping probability is $p$, because $\rho_{A_1B_1}$ is transformed as $\Lambda(\rho_{{A_1}{B_1}}) = d^2\hbox{tr}_{{A_2}{B_2}}E_{{A_1}{A_2}{B_1}{B_2}}(I\otimes \rho_{A_2B_2}^T)$.
\end{itemize}
The optimal swapping probability under PPT operations is then obtained by maximizing $p$ under the constrains (a)-(d).
This is a semidefinite programming problem \cite{Boyd04a}.
We numerically calculated the optimal probability $p$ for random density matrices in two qubits ($d=2$) as follows.
First, we generated the eigenvalues of a density matrix $\{ \lambda_1,\lambda_2,\lambda_3,\lambda_4\}$ uniformly distributed under $\lambda_1+\lambda_2+\lambda_3+\lambda_4=1$ (in the case where the rank of the density matrix is 4) \cite{lambda}. 
Second, we generated a $4\times 4$ random unitary matrix according to \cite{uni}, where a numerical method to generate random unitary matrices with the Haar measure (thus representative for circular unitary ensemble) is provided. We then obtained a random density matrix by applying the unitary transformation to the diagonal matrix whose diagonal elements are $\{ \lambda_1,\lambda_2,\lambda_3,\lambda_4\}$. 
Finally, we solved the semidefinite programming problem for this density matrix numerically using the computer code \cite{sdp} to find the optimal $p$. 

\begin{figure}[tb]
\centerline{\epsfig{file=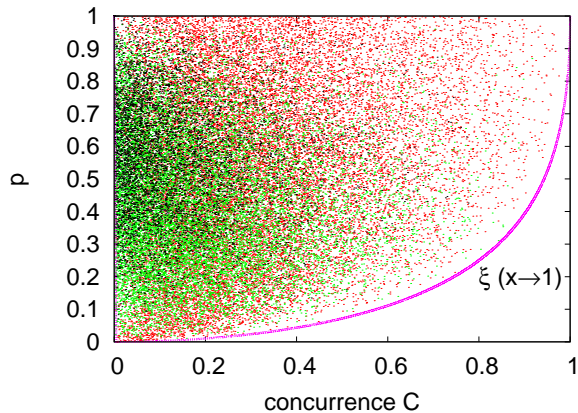, width=8.2cm}} 
\vspace*{13pt}
\caption{The relationship between the swapping probability $p$ and the concurrence $C$. Red, black, and green dots represent the result for rank 2, 3 and 4, respectively.}
\end{figure}

\section{Numerical results}
\label{s4}
Figure 2 shows the relationship between the optimal swapping probability $p$ and the concurrence $C$. 
The results for 20,000 random density matrices are plotted in each case of rank 2, 3, and 4. 
From this figure, it turns out that $p$ is bounded below by a nonzero probability.
This implies that almost all states can be swapped with a non-vanishing probability by PPT operations.
Moreover, it turns out from the figure that the lower bound of $p$ increase as $C$ increases. 
In order to investigate this lower bound, let us consider the optimal swapping probability $p_\xi$ of the mixed state $\xi$.
As mentioned in Sec. \ref{s2}, $p_\xi$ is independent on $x$ but the concurrence of $\xi$ becomes maximum when $x\to 1$, because $\xi$ approaches to the pure state of $\ket{\psi}$ in this limit.
Namely, when $x\to 1$, $\xi$ has as high entanglement as the pure state, but the swapping probability remains $y/(1-y)$.  
The curve in Fig. 2 shows the relationship between $p_\xi$ and $C_\xi$ in the limit of $x\to 1$, and our numerical results strongly suggest that the swapping probability is lower bounded by this curve.
It is quit interesting that this is an open bound because the curve is reached only when $x \to 1$.
Note that a similar result is obtained for the other entanglement measure of negativity.
 
\begin{figure}[tb]
\centerline{\epsfig{file=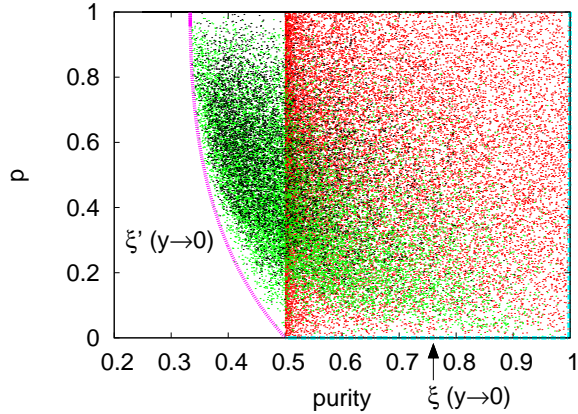, width=8.2cm}} 
\vspace*{13pt}
\caption{The same as Fig. 2, but the relationship between the swapping probability $p$ and the purity.}
\end{figure}

Figure 3 shows the relationship between the optimal swapping probability $p$ and the purity.
Note that, when the purity is less than or equal to 1/3, there are no entangled states \cite{lambda}, 
and hence $p = 1$ in this region.
When the purity is greater than or equal to 1/2, $p$ distributes from $0$ to $1$, and there is no nonzero lower bound.
This is because the state $\xi$, whose purity is determined by $x$ only and takes a value in [1/2, 1), gives a vanishing swapping probability in the limit of $y \to 0$. 
When the purity is less than 1/2, 
however, the state $\xi$ is absent because the states of rank 2, including $\xi$, cannot take the purity less than 1/2.
In this region, it is found from the figure that $p$ has a nonzero lower bound.
This fact also supports that the swapping probability of $\xi$ is particularly low because $p$ is considerably raised by the absence of $\xi$.
In this region, we found that the lower bound is given by the following state $\xi'$
\begin{equation}
\xi' = x'(\ket{\psi}\bra{\psi}+\ket{01}\bra{01}) + (1-2x')\ket{\psi_{\perp}}\bra{\psi_{\perp}},
\end{equation}
where $\frac{1}{3}< x' < \frac{1}{2}$, and $\ket{\psi_{\perp}} = \sqrt{1-y}\ket{00} - \sqrt{y}\ket{11}$.
Indeed, the numerically obtained swapping probability of $\xi'$ in the limit of $y \to 0$ shown by a curve in Fig. 3 clearly lower bounds all the data points for random density matrices.
Note that the state $\xi'$ coincides with $\xi$ at $x' = \frac{1}{2}$.

\begin{figure} [tb]
\centerline{\epsfig{file=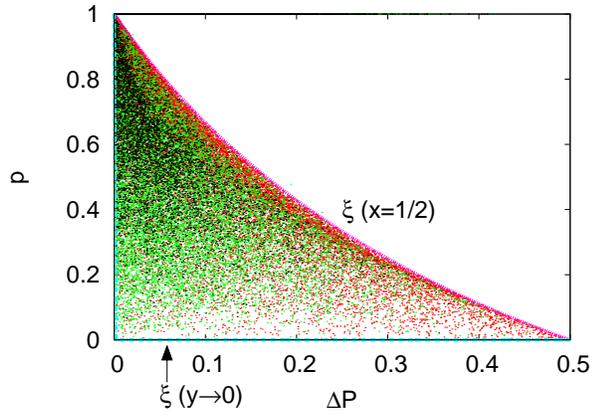, width=8.2cm}} 
\vspace*{13pt}
\caption{The relationship of swapping probability $p$ and $\Delta P$.}
\end{figure}

Figure 4 shows the relationship between the swapping probability $p$ and the absolute value of the difference of the purity of the reduced density matrices ($\Delta P$).	
Only when $\Delta P = 0$, the optimal swapping probability reaches 1, which reproduces the result by Horodecki {\it et. al.} \cite{KMP}.
As in the case of Fig. 3, there is no nonzero lower bound, which is again explained by the optimal swapping probability of $\xi$ in the limit of $y \to 0$, because $\Delta P_{\xi}$ can take any value within the range of $0 <\Delta P_\xi< \frac{1}{2}$ [see Eq. (\ref{dp})].
Figure 4 also shows that the upper bound of $p$ exists and decreases with the increase of $\Delta P$.
The swapping probabilities of the states $\xi$ with $x = \frac{1}{2}$ and $0<y\leq \frac{1}{2}$ upper bound all the data points (except for a few points of rank 3, which probably originate from the instability of the numerical calculations).

\section{Summary and discussion}
\label{s5}
In this paper, we performed numerical calculations to study how the optimal probability of swapping a state by PPT operations is related to the amount of the entanglement (concurrence and negativity).
As a result, we numerically showed that almost all states in two qubits can be swapped with nonzero probability, and the lower bound of the optimal swapping probability increases for increasing the entanglement.
In particular, we showed that the lower bound approaches to zero in the limit of vanishing entanglement.
This is somewhat surprising because unentangled states can always be swapped with a unit probability.
Moreover, numerical results strongly suggest that the lower bond corresponds to the swapping probability of the state $\xi$ in the limit of $x \to 1$.
Namely, this state is the most difficult state to swap among the states which have the same amount of entanglement in two qubits.

Why is the state $\xi$ extremely difficult to swap?
The reason is not so clear but this state, the mixture of the pure entangled states $\ket{\psi}$ and $\ket{01}$, has relatively high entanglement. 
In order to swap this state, however,
we are subject to the strong constraint that $\ket{01}$ must be transformed to $\ket{10}$, even if the portion of $\ket{01}$ is infinitesimally small (see Appendix A).
As a result, $\xi$ has the special property that the optimal swapping probability is fully discontinuous around $x = y = 1$,
i.e. the probability drops from 1 to 0 by the infinitesimal change of $x$. 

\begin{figure} [tb]
\centerline{\epsfig{file=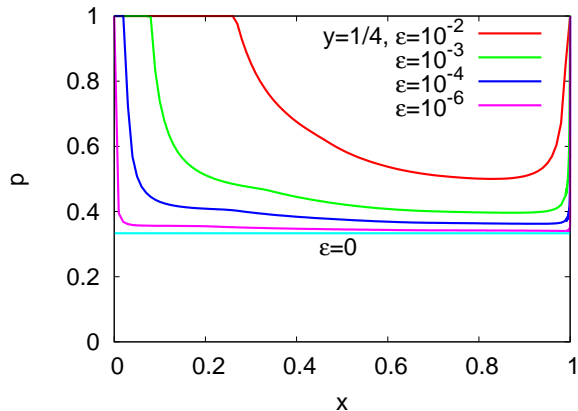, width=8.2cm}} 
\vspace*{13pt}
\caption{The optimal swapping probability of $\xi_{AB}$ with y=1/4 for various values of $\epsilon$.}
\end{figure}

Obviously, this discontinuity originates from the fact that we considered the exact transformation.
If we consider the approximate transformation where the target state $\rho_{AB}'$ is allowed to have a small margin $\epsilon$ such that $D(\rho_{AB}',V\rho_{AB} V) \leq \epsilon$ ($D$ is the trace distance), the discontinuity is smeared out.
Indeed, in this approximate scheme, we have to do nothing to swap for $\xi$ with $1-x \leq \epsilon$, hence $p=1$ and the discontinuity at $x=1$ is resolved.
However, even in this case, $p$ again approaches to $y/(1-y)$ for $1-x \sim \epsilon$ as shown in Fig. 5,
and the concurrence is still as high as $\sim(1-\epsilon)\sqrt{y(1-y)}$.
Therefore, the curve of Fig. 2 is only slightly shifted to the left (for small $\epsilon$) and our conclusion is not changed critically.

We also studied how the swapping probability is related to the mixedness, and showed that the lower bound ($p = 0$ for a purity larger than 1/2) is again explained by the state $\xi$ (in the limit of $y \to 0$).
Therefore, $\xi$ is the most difficult state to swap in this sense also. 

We hope that our results will shed some light on unknown properties of entanglement, in particular concerning the asymmetry of quantum correlation.  

\begin{acknowledgments}
This work was supported by JSPS KAKENHI Grants No. 23246071 and No.24540405. 
\end{acknowledgments}

\appendix

\section{}

In this appendix, we prove that $p_{\xi}$ is given by $y/(1-y)$.
An SLOCC map $\Lambda$ has the following properties of ($\alpha$)-($\gamma$):
\begin{itemize}
\item[($\alpha$)] $\Lambda(\rho) = \sum_i (A_i \otimes B_i)\rho (A_i \otimes B_i)^\dagger$ \ (the Kraus representation),

\item[($\beta$)] A pure state is transformed to a pure state for each $i$ (i.e. $(A_i \otimes B_i)\ket{\chi_{AB}}$ is a pure state),

\item[($\gamma$)] An unentangled state cannot be transformed to an entangled state.
\end{itemize}
Now, suppose that $\Lambda$ achieves the swap of $\xi \to pV\xi V$ with $p$ being the swapping probability, and therefore 
\begin{eqnarray}
\Lambda (\xi )&=&x\Lambda (\ket{\psi}\bra{\psi})+(1-x)\Lambda (\ket{01}\bra{01}) \nonumber \\
&=&px\ket{\psi}\bra{\psi} + p(1-x)\ket{10}\bra{10}.
\end{eqnarray}
From the property of ($\beta$) and the above, it is found that $(A_i\otimes B_i)\ket{01}$ must be a pure state which belongs to the space spanned by $\ket{\psi}$ and $\ket{10}$. 
On the other hand, the pure state must be an unentangled state from the property of ($\gamma$).
The pure state $a\ket{\psi}+b\ket{10}$, however, is not entangled only when $a=0$ (namely, there is only an unentangled state $\ket{10}$ in the space),
and as a result, $(A_i \otimes B_i)\ket{01}\propto \ket{10}$ must be fulfilled for all $i$.
Therefore, we have
\begin{eqnarray}
\Lambda (\ket{01}\bra{01} )    &=&q\ket{10}\bra{10},\\
\Lambda (\ket{\psi}\bra{\psi} )&=&p\ket{\psi}\bra{\psi} + (p-q)\frac{1-x}{x}\ket{10}\bra{10}.
\end{eqnarray}
Let us then consider the Kraus operator $A_i$ and $B_i$ by components:
\begin{eqnarray}
A_i &=& a_{00}^{(i)}\ket{0}\bra{0}+a_{10}^{(i)}\ket{1}\bra{0}+a_{01}^{(i)}\ket{0}\bra{1}+a_{11}^{(i)}\ket{1}\bra{1}, \nonumber \\ \label{kousatu1} \label{Ai}\\
B_i &=& b_{00}^{(i)}\ket{0}\bra{0}+b_{10}^{(i)}\ket{1}\bra{0}+b_{01}^{(i)}\ket{0}\bra{1}+b_{11}^{(i)}\ket{1}\bra{1}. \nonumber \\ \label{kousatu2} \label{Bi}
\end{eqnarray}
From $(A_i \otimes B_i)\ket{01} \propto \ket{10}$, we have $a_{00}^{(i)}b_{11}^{(i)} = 0$.
If $a_{00}^{(i)} \neq 0$ and $b_{11}^{(i)} =0$ are assumed, we have $a_{00}^{(i)}b_{10}^{(i)} = 0$ from the fact that $A_i \otimes B_i \ket{\psi}$ does not have the component of $\ket{01}$, and hence  $b_{10}^{(i)} = 0$, which implies that $\braket{11|A_i \otimes B_i|\psi} = 0$ and the swapping cannot be achieved.
Therefore $a_{00}^{(i)} = 0$ must be fulfilled. Similarly, $b_{11}^{(i)} = 0$ must be also fulfilled. Then,
\begin{equation}
(A_i \otimes B_i)\ket{\psi}=(a_{10}^{(i)}b_{00}^{(i)}\sqrt{y}+a_{11}^{(i)}b_{01}^{(i)}\sqrt{1-y})\ket{10}+a_{10}^{(i)}b_{10}^{(i)}\sqrt{y}\ket{11}+a_{01}^{(i)}b_{01}^{(i)}\sqrt{1-y}\ket{00}. 
\end{equation}
In order that this is a superposition of $\ket{\psi}$ and $\ket{10}$, we have
\begin{equation}
a_{10}^{(i)}b_{10}^{(i)}\sqrt{\frac{y}{1-y}} = a_{01}^{(i)}b_{01}^{(i)}\sqrt{\frac{1-y}{y}}.
\end{equation}
Therefore,
\begin{equation}
(A_i \otimes B_i)\ket{\psi} = (a_{10}^{(i)}b_{00}^{(i)}\sqrt{y}+a_{11}^{(i)}b_{01}^{(i)}\sqrt{1-y})\ket{10}\nonumber+a_{10}^{(i)}b_{10}^{(i)}\sqrt{\frac{y}{1-y}}\ket{\psi}.
\end{equation}
From this, since $p$ is the coefficient of $\ket{\psi}\bra{\psi}$ in $\Lambda (\ket{\psi}\bra{\psi})$,
we have
\begin{equation}
p=\frac{y}{1-y}\sum_i |a_{10}^{(i)}b_{10}^{(i)}|^2 \label{pmax}.
\end{equation}
On the other hand, since $A_i \otimes B_i$ constitutes a part of positive operator valued measure (POVM),
$\sum_i A_i^\dagger A_i \otimes B_i^\dagger B_i \leq I$
must be fulfilled.
Therefore, we have 
$\sum_i |a_{10}^{(i)}|^2(|b_{10}^{(i)}|^2 + |b_{00}^{(i)}|^2) \leq 1 \label{leqleq}$,
and consequently
\begin{equation}
p = \frac{y}{1-y}\sum_i |a_{10}^{(i)}|^2 |b_{10}^{(i)}|^2 \leq \frac{y}{1-y} \label{pmax_ika}.
\end{equation}
This upper bound is indeed reached by the Kraus operator of Eq. (\ref{kkk}).

\end{document}